\documentstyle[prl,aps,multicol,epsfig]{revtex}
\begin{document}

\draft

\title{
A phase-field model of Hele-Shaw flows in the high viscosity contrast regime 
}

\author{A. Hern\'{a}ndez-Machado$^1$, A.\ M. Lacasta$^2$, E. Mayoral$^3$ and 
E. Corvera Poir\'e$^3$\\}
\address{
$^1$ Departament ECM, Facultat de F\'{\i}sica, Universitat de
Barcelona\\
Diagonal 647, E-08028 Barcelona, Spain \\
$^2$ Departament de F\'{\i}sica Aplicada,
Universitat Polit\`{e}cnica de Catalunya\\
Avda. Dr. Mara\~{n}on 44, E-08028 Barcelona, Spain.\\    
$^3$ Departamento de F\'{\i}sica y Qu\'{\i}mica Te\'orica, Facultad de 
Qu\'{\i}mica, UNAM\\
Ciudad Universitaria, M\'exico, D.F. 04510, M\'exico} 
\date{\today}
\maketitle

\begin{abstract}
A one-sided phase-field model is proposed to study the dynamics of unstable
interfaces of  Hele-Shaw flows in the high viscosity contrast regime. The corresponding macroscopic equations are obtained 
by means
of an asymptotic expansion from
the phase-field model. Numerical integrations of the phase-field model
in a rectangular Hele-Shaw cell 
reproduce finger competition with the final evolution to a 
steady state 
finger.
\end{abstract}

\medskip
\maketitle


\begin{multicols}{2}
\narrowtext

The characterization of the dynamics of morphologically unstable 
interfaces is 
one of major problems of non-equilibrium phenomenology \cite{general}.
Some relevant examples of interfaces that grow out of equilibrium are
dendritic growth, directional solidification, flow in porous media,
electro-deposition, bacterial colony growth and two-fluid flow
in a Hele-Shaw cell. The latter example is also called the Saffman-Taylor
problem and has played a central role in this field,
both because of its relative simplicity and because of its potential
importance in oil recovery. It has been widely studied both
experimentally and theoretically.

Even if the Saffman-Taylor problem is mathematically simple in relation to other
problems,
it has a moving boundary condition which makes it a free-boundary problem.
The corresponding equations have been solved
analytically for very short times by means of a linear stability analysis
and for the steady state finger shape by means of conformal mapping 
techniques \cite{st,mclean}. Some analytical results have also been
obtained for the dynamics of intermediate time \cite{tanveer}.
Numerically there are several techniques, most of them involving
integral boundary methods \cite{aref,integro,hou}

The so-called phase-field models have been introduced within the 
context of solidification to study the dynamics from the linear regime to 
the long time behaviour \cite{phase}. 
These models are based on the 
introduction of a mesoscopic equation for an order parameter (the phase-field).
This equation is coupled to other physical fields (such as a thermal field). 
The advantage of this method is that
one does not have to explicitly trace the interface. It is a field model
for all values of the order parameter that varies continuously from
one phase to the other. One has to identify the locus of points
with a given value of the order parameter, which is arbitrarily chosen
to be the interface. 
The use of a mesoscopic model, for which the interface has a small
width $\epsilon$, is justified as long as in the sharp interface limit, 
$\epsilon \to 0$, the correct macroscopic equations are recovered.
Recently, the concept of phase-field models has been used in a broader sense
to include any model which contains continuous fields that are introduced
to describe phases separated by diffuse interfaces.
Phase-field models have been used in a wide range of problems
such as viscous fingering, roughening, vesicles, pinchoff and reconnection in a Hele-Shaw
cell and intracellular dynamics 
\cite{folch,gonzalez,machado,misbah,lowengrub,levine}.

In general, the phase-field models have been considered for symmetric 
situations where 
the characteristic parameters (such as the thermal diffusivity) are 
identical in both phases. This gives rise to the so-called two-sided 
symmetric models. 
Very recently, Karma \cite{karma} has proposed a phase-field model 
of the one-sided type (with zero diffusion in one phase)
to simulate quantitatively micro-structural pattern formation of
alloy solidification.
For the viscous fingering problem with arbitrary viscosity contrast,
a phase-field model has been introduced in Refs. \cite{folch}.
Such phase-field model, which is a two-sided model, is useful to describe
the problem of viscous fingering except in the high viscosity
contrast regime.
This regime is experimentally relevant since typically the pushing fluid
is air or other fluid of negligible viscosity.
For such a regime, a proper model was laking and
this is what we are presenting in this paper:
a one-sided phase field model for the high viscosity contrast
regime of the viscous fingering problem.

Our model contains an equation for an order parameter. 
It is Model B of Ginzburg-Landau phenomenology \cite{hohenberg}.
Instead of the 
coupling of the order parameter to a physical field  through a second 
equation, we include a
boundary condition such that the interface becomes unstable. This is done by means of a ramp that creates a flux from
the boundary. To consider a one-sided model, we only need to 
neglect changes in the order parameter in one of the two phases. The model 
could also be relevant for dendritic growth at very small undercooling
by introducing anisotropy
\cite{almgren,guo}.  

Our phase-field model has the advantage
of being very simple to be implemented on a computer and contains a complete description of all the nonlinear and nonlocal properties of the macroscopic model. 
We show how the macroscopic equations of the problem
are obtained from the phase-field model in the sharp interface limit.
This is done by means of the matched  asymptotic expansion method.
We then present numerical solutions showing how our phase-field model 
reproduces the main features of the viscous
fingering problem such as the dynamic competition of modes and
the formation of a steady state finger.
This makes the model an attractive tool to be used to study problems
that would not be easily feasible with traditional methods such
as the propagation of viscous fingering in the presence of
quenched disorder.

\section*{The model}
\subsection*{The Viscous Fingering Problem}
In the Saffman Taylor problem 
both fluids are governed by Darcy's law, which relates the fluid velocity
to the pressure gradient \cite{st}. 
When the low viscosity fluid displaces the high viscosity fluid, the interface
between both fluids is unstable. 
When the pushing fluid is considered to have zero viscosity,
Darcy's law states that the pressure on the pushing fluid is constant
and all that remains to be solved are the equations for the viscous fluid
subject to the proper boundary conditions at the fluid-fluid interface.
This is called the high viscosity contrast regime and in this regime
the equations for the displaced viscous fluid are 

\begin{equation}
\label{presion}
\nabla^2 p=0,
\end{equation}                                                                     

\begin{equation}
\label{conti}
v_n= -K\nabla p \cdot \hat{n},
\end{equation}                                                                     
\begin{equation}
\label{gibbs}
\Delta p=\gamma \kappa.
\end{equation}   
Eq.\ (\ref{presion}) is the Laplace equation in the bulk,
where $p$ is the pressure of the viscous fluid.
At the interface, there are two
boundary conditions: the continuity equation, Eq.\ (\ref{conti}) and the 
Gibbs-Thomson condition, Eq.\ (\ref{gibbs}).
$v_n$ is the velocity normal to the interface. $K$ is the permeability of the 
viscous fluid, 
$K=\frac{b^2}{12\mu}$, where $b$ is the separation
between the plates and
$\mu$ is the viscosity of the fluid that is being pushed.
$\Delta p$ is the pressure of the viscous fluid
 minus the
constant pressure at the zero viscosity fluid, which without loss
of generality can be taken equal to zero. 
$\kappa$ is the local
curvature at the interface and $\gamma$ is the surface tension.
These three equations also describe solidification in the quasi-static
limit of small undercooling by introducing anisotropy.
In what follows, we present the equations for our phase-field model and show how
it reproduces the above equations in the sharp interface limit. 

\subsection*{Phase-field model}

Our phase-field model 
contains a time-dependent Ginzburg-Landau equation for a conserved order
parameter and includes a boundary
condition that makes the interface unstable.
The equation reads
\begin{equation}
\label{modelb}
\nonumber
\frac{\partial\phi}{\partial t}=\nabla 
\cdot \left [ M_0
\nabla(-\phi+\phi^3 - \epsilon^2\nabla^2\phi) \right ] .
\end{equation}
The local order parameter $\phi$ adopts the equilibrium values
$\phi_{eq}=1$ (air-phase) and $\phi_{eq}=-1$ (viscous fluid-phase).
At the interface, $\phi$
varies continuously from one phase to the other.
The parameter $M_0$ has a constant value
in each phase and is zero in air

\begin{eqnarray}
\label{mobility}
M_0=
\left \{
\begin{array}{ll}
M, &  \rm{if\ } {\it \phi < 0}, \\ 
m=0 &  \rm{if\ } {\it \phi \geq 0}.\\ 
\end{array} 
\right .
\end{eqnarray} 

\begin{figure}
\begin{center}
\def\epsfsize#1#2{0.50\textwidth}
\leavevmode
\epsffile{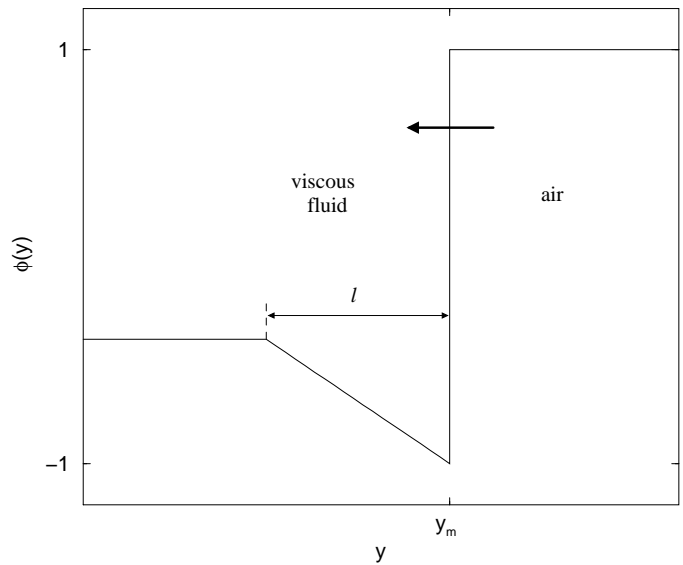}
\caption{
Scheme of the initial profile prepared with a ramp that will be
maintained during the temporal evolution.
}
\label{fig1}
\end{center}
\end{figure}
The air phase can be pulled toward the viscous fluid. 
An unstable interface is developed by
maintaining a slope in the order parameter close to the interface,
as the case shown in Fig.\ \ref{fig1}. 
This situation can be created by initially preparing the system 
with a profile of the form 
\begin{eqnarray}
\label{ic}
\phi(x,y)=
\left \{
\begin{array}{ll}
1&  \rm{if\ } {\it y > y_m},\\
-1-\alpha(y-y_m)&  \rm{if\ } {\it y_m-l < y \leq y_m},\\
-1+ \alpha l&  \rm{if\ } {\it y \leq y_m-l},\\
\end{array}
\right .
\end{eqnarray}
and fixing the value $\phi_f=-1+\alpha l$ behind the interface up to a distance 
$l$ throughout all the temporal evolution. 
This slope represents the driving force of the system.
The parameter $\alpha$ controls the slope and the fingers growth
velocity.
For convenience we will refer to the air as the {\it plus} phase
and to the viscous fluid as the {\it minus} phase.

\section*{The sharp interface limit}

In this section we obtain the macroscopic equations for the viscous fingering
problem in the high viscosity contrast regime by means of an
asymptotic expansion of the phase-field model in the sharp interface limit
$\epsilon \rightarrow 0$ \cite{foot}.
   
Our starting point is Eq.\ (\ref{modelb}) used in
the study of a conserved order parameter,
$\phi$.
The chemical potential is given by 
\begin{equation}
\mu (\phi)=\mu_B -\epsilon^2\nabla^2\phi=-\phi+\phi^3
-\epsilon^2\nabla^2\phi\label{213}.
\end{equation}

We divide the space into an outer
and an inner region. 
We assume that $\phi=\pm\phi_{eq} + O(\epsilon)$ far from the interface. 
$\epsilon$ is considered to be a small parameter and we expand all the variables 
$a({\bf r},t)$ around the value $\epsilon=0$
in the outer region. We obtain

\begin{equation}
a({\bf r},t)=a_0 +\epsilon a_1+\epsilon^2 a_2 + ... \label{215}
\end{equation}

For the interfacial region or inner region, we adapt our coordinate 
system using time dependent curvilinear coordinates.
The interfacial points  are given by the curvilinear coordinates
$u$, which is the normal distance to the interface, and $s$, which
is the arclength. 
Because the natural dimension in the inner region must be small ,
we introduce the variable $w$ defined as $w=\frac{u}{\epsilon}$.
Thus, in the sharp interface limit, when $\epsilon \rightarrow 0$, 
the inner region goes from $w\to -\infty$ to $w\to +\infty$. 
We use the corresponding inner fields $A(w,s,t)$ in the inner region and
the corresponding expansion is

\begin{equation}
A(w,s,t)=A_0 +\epsilon A_1+\epsilon^2 A_2 + ... \label{216}
\end{equation}

When we take the limit of the sharp interface, $\epsilon\to 0$,
the conditions for the field $a$ and $A$ ,
from the expansions to $i$-th order in $\epsilon$ are

 \begin{equation}
 \lim_{w \to -\infty}A_i=\lim_{u \to -0}a_i\label{226},
 \end{equation}

 \begin{equation}
 \lim_{w\to -\infty}\partial_wA_{i+1}=\lim_{u\to -0}\partial_ua_i\label{227}.
 \end{equation}
Due to the fact that $m=0$ in air, the 
matching condition is only imposed in the viscous phase.
 
In the inner region, we introduce the order parameter
$\tilde\phi$ such that $\tilde\phi(u(t),s,t)=\phi(\bf r,t)$,
therefore

 \begin{equation}
 \partial_t\phi=\partial_t \tilde\phi+\partial_t u \partial_u \tilde\phi\label{221}.
 \end{equation}

We rescale time as $\tau=\epsilon t$ 
since we work in the quasi-static approximation, where the
characteristic times for interface motion are much larger than the characteristic times
for the diffusion to take place.
The local curvature $\kappa=-\nabla^2 u$ is positive 
when a bump of the $\phi>0$ phase protrudes into the $\phi<0$ phase.
Starting from Eq.\ (\ref{221}), using the Laplacian operator
in curvilinear coordinates 
$ \nabla^2={\partial_u}^2-\kappa\partial_u+\partial_s^2$,
and doing the corresponding variable changes, we have

 \begin{eqnarray}
\epsilon\partial_\tau\phi-\frac{v}{\epsilon}\partial_w\phi &=M(&\frac{1}{\epsilon^2}{\partial_w}^2\mu(\phi)-\frac{\kappa}{\epsilon}\partial_w \mu(\phi) +{\partial_s}^2\mu(\phi))\label{223},
 \end{eqnarray}                                                                     
where we have dropped the tildes. The normal velocity 
$v=-\partial_t u $ is positive if the phase with a negative
order parameter goes into the phase with a positive order parameter. 
This variable is also expanded in powers of $\epsilon$.

For the chemical potential $\mu$, the inner expansion in terms
of $\phi$ (to order $\epsilon^2$) is given by

\begin{eqnarray}
\mu(\phi)=\mu_0+\epsilon \mu_1 +\epsilon^2 \mu_2,
\label{224}
\end{eqnarray}
with
 \begin{eqnarray}
 \mu_0&=&\mu_{B_0}-\partial^2_w\phi_0, \nonumber \\ 
 \mu_1&=&\mu'_{B_0}\phi_1-\partial^2_w\phi_1+\kappa\partial_w\phi_0,
\label{224b}\\
 \mu_2&=&\frac{1}{2}\mu''_{B_0}(\phi_1)^2+\mu'_{B_0}\phi_2-\partial_w^2\phi_2+\kappa\partial_w\phi_1-\partial_s^2\phi_0,
\nonumber
\end{eqnarray}
where $\mu_{B_0}=\mu_{B}(\phi_0)$. The prime represents the derivative of $\phi$ evaluated at $\phi_0$.

 For the region far from the interface (outer region), the length scale 
involved is much greater than $\epsilon$, so we can use a common 
time-independent coordinate system.
 In the viscous fluid region the dynamical equation for the order parameter 
is simply

 \begin{equation}
 \epsilon\partial_\tau\phi=M\nabla^2\mu(\phi)\label{225},
 \end{equation}
where $\mu(\phi)$ is given by Eq.\ (\ref{213}).

\subsection*{Inner region}

We now proceed to solve the equations for the inner region
Eqs.\ (\ref{223})-(\ref{224b}). 
We also use the matching conditions Eq.\ (\ref{226})
and Eq.\ (\ref{227}). Solutions that obey  $\phi(0)=0$ and
$\partial_w\phi_0(-\infty)=0$ are required.   

{\bf order $\epsilon^{-2}$}    For the inner region, the dynamical equation to lowest order in $\epsilon$,
($\epsilon^{-2}$) is taken from Eq.\ (\ref{223})

 \begin{equation}
 \partial^2_w \mu_0=0\label{231}.
 \end{equation}
Here we have taken into account the expansion for $\mu$.
The previous expression has a solution  $\mu_0=m_0+n_0w$. 
The requirement that $\mu_0$ must be finite for $w \to -\infty$ implies
that $n_0=0$.
Finally, we consider $m_0=0$ and then $\phi_{eq}=\pm 1$.
Therefore, $\mu_0=0$ in the inner region.

{\bf order $\epsilon^{-1}$} Taking the first-order terms $\epsilon^{-1}$ 
from the dynamical equation in the inner region, Eq.(\ref{223}), we have

 \begin{equation}
 -v_0\partial_w\phi_0=M\partial_w^2\mu_1\label{233},
 \end{equation}
since $\mu_0=0$.
Integrating Eq.\ (\ref{233}) in $w$  we find

 \begin{equation}
 -v_0\phi_0=M\partial_w\mu_1+n_1\label{235}.
 \end{equation}

By evaluating 
 between the limits $w=-\infty$ and $w=\infty$
 Eq.\ (\ref{235}), we obtain

 \begin{equation}
 -2 \phi_{eq} v_0=M\partial_w\mu_1(-\infty)\label{236},
 \end{equation}
where  $2\phi_{eq}$ is the order parameter change between 
the two phases and we have only the contribution of the viscous fluid phase
on the right-hand side.
Using  Eq.\ (\ref{227}) we have 
 $\partial_w\mu_1(-\infty) =\partial_u\mu_0(-0)=0$.
>From Eq.\ (\ref{236}),
$v_0$ also vanishes and 
Eq.\ (\ref{235}) gives  $n_1=0$.

By integrating Eq.\ (\ref{235})
in $w$, we find that 
$\mu_1=\mu_1(-\infty)$ is a constant. 
In order to obtain $\mu_1(-\infty)$, we use its expression from Eq.\ (\ref{224b}) and 
we multiply both sides of this expression 
by $\partial_w\phi_0$ and integrate in $w$

 \begin{eqnarray}
 \mu_1(-\infty)\int dw\partial_w\phi_0&=&
\int dw\partial_w\phi_0(\mu'_{B_0}-\partial^2_w)\phi_1  
\nonumber \\
&+&\kappa\int dw (\partial_w\phi_0)^2\label{238}.
 \end{eqnarray}
The function  $\partial_w\phi_0$ is known as the  Goldstone mode
and is related to the translational invariance of the interface. 
The equation for  $\phi_0$, written as a function of the rescaled
variable $w$, is $\mu_{B_0}-\partial^2_w\phi_0=0$. Differentiating
with respect to $w$ we obtain an equation for $\partial_w\phi_0$, which is
 $(\mu'_{B_0}-\partial^2_w)\partial_w\phi_0=0$. So the Goldstone mode
is a zero eigenvector of the linear operator $\mu'_{B_0}-\partial^2_w$.
By doing integration by parts, the
first term on the right-hand side of Eq.\ (\ref{238})
vanishes and we obtain

 \begin{equation}
\mu_1(-0)=\frac{\gamma}{\phi_{eq}}\kappa\label{2313},
 \end{equation} 
where $\gamma=\frac{1}{2}\int dw (\partial_w\phi_0)^2$ 
is the surface tension and
we have used 
 the matching condition for $\mu_1(-0)$ from  Eq.\ (\ref{226}).
Taking into account the fact that at the interface $p(-0)=\phi_{eq}\mu_1(-0)$ we obtain 
Eq.\ (\ref{gibbs}).

{\bf order $\epsilon^{0}$} In order to obtain 
the continuity equation,
we need to go to the next order.
The dynamical equation to order  $\epsilon^0$ in the inner region is

 \begin{equation}
 -v_1\partial_w\phi_0=M({\partial_w}^2\mu_2-\kappa\partial_w\mu_1+{\partial_s}^2\mu_0)\label{2314}.
 \end{equation}

Integrating Eq.\ (\ref{2314})
in the direction normal to $w$ we find
\begin{equation}
-2 \phi_{eq} v_1=M\partial_w\mu_2(-\infty)=M\partial_u\mu_1(-0)\label{2316},
\end{equation}
where we have used the matching condition Eq.\ (\ref{226}), and the fact
that $\partial_w\mu_1(-\infty)=0$ and that $\mu_0=0$.
Eq.\ (\ref{2316}) could be written as Eq.\ (\ref{conti}) in terms of 
the pressure
at the interface of the viscous fluid, where $K=M/(2\phi_{eq}^2)$.

\subsection*{Outer region}  

{\bf order $\epsilon^{-2}$} The dynamical equation in the outer region
to lowest order $\epsilon^{-2}$ is
 
 \begin{equation}
 \nabla^2\mu_0=0\label{232}.
 \end{equation}
The boundary condition far from the interface is then $\mu_0=0$.
We previously found that $\mu_0=0$ for the inner solution at the interface.
The only solution satisfying both conditions is $\mu_0=0$. It follows that
$\phi_0=\phi_{eq}$ in the \emph{plus} phase and $\phi_0=-\phi_{eq}$
in the \emph{minus} phase.
This was to be expected since the lowest order in the expansion
corresponds to the solution of the flat interface.

{\bf order $\epsilon^{-1}$} The dynamical equation for the order $\epsilon^{-1}$ is
 
\begin{equation}
\label{laplace}
\nabla^2\mu_1=0\label{2310}.
\end{equation}
At order $\epsilon$ the order parameter and the chemical potential are 
proportional and from Eq.\ (\ref{laplace}) we obtain  
Eq.\ (\ref{presion}). 

\section*{Numerical Results}

We have numerically integrated Eqs.\ (\ref{modelb})-(\ref{mobility})
with $\epsilon=1$ and $M=1$ on a rectangular lattice of
vertical size $\L_y=200$ and mesh size $\Delta x=1$,
with periodic boundary conditions in the $x$ direction and reflecting
boundary conditions in the $y$-direction. 
The system has been prepared with a horizontal
interface containing some perturbations in order to be destabilized. 
The profile in the vertical direction 
is done by Eq.\ (\ref{ic}) with $l=10$. 
As was mentioned before, during the evolution we maintain 
a slope by fixing the value 
$\phi_f=-1+\alpha l$ behind the interface,
up to a distance $l$ measured from the tip of the most advanced
finger.

\subsection*{Finger competition}
\begin{figure}
\begin{center}
\def\epsfsize#1#2{0.55\textwidth}
\leavevmode
\epsffile{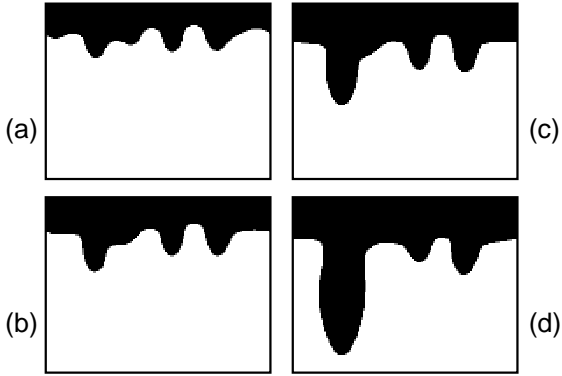}
\caption{
Finger development and competition for $\alpha=0.04$
corresponding to early times:
$t=100$ (a), $500$ (b), $1000$ (c) and $2000$ (d).
}
\label{fig2}
\end{center}

In order to better visualize the competition process
we have prepared a second initial condition consisting of
two well-formed fingers, in which one of them is a bit larger than
the other. In Fig.\ref{fig3} we observe how the largest
finger grows at the expense of the other, which
moves backwards becoming smaller and  
eventually  disappears.

\end{figure}

Firstly, we are interested in the generation and 
subsequent competition of fingers during the
early stages of the evolution.
A wide system of size $L_x=128$ has been considered
and we have prepared an
initial corrugated interface 
formed by the superposition of several modes of random
amplitude.
In Fig. \ref{fig2} we show a typical evolution.
It is seen that fingers develop from the random initial configuration.
Some modes grow, some modes
decay and  finger competition begins. Both features have been observed
in theoretical and experimental studies of the viscous fingering problem.
The competition process continues until only one of the 
fingers survives.
\begin{figure}
\begin{center}
\def\epsfsize#1#2{0.60\textwidth}
\leavevmode
\epsffile{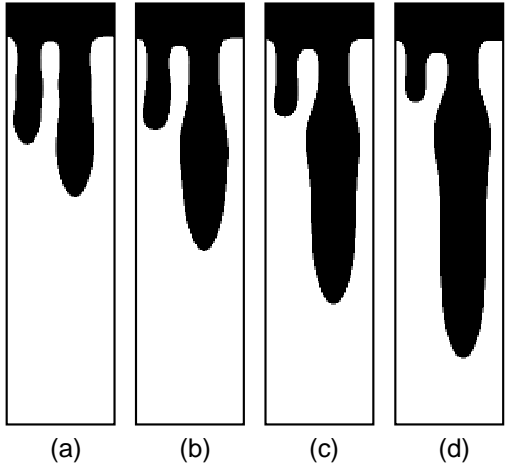}
\caption{
Finger competition process for two initially well-formed
fingers with
$\alpha=0.04$ and system width $\L_x=64$.
The patterns are  separated by time intervals of $1000$.
}
\label{fig3}
\end{center}
\end{figure}
\subsection*{Steady-state finger}
\begin{figure}
\begin{center}
\def\epsfsize#1#2{0.50\textwidth}
\leavevmode
\epsffile{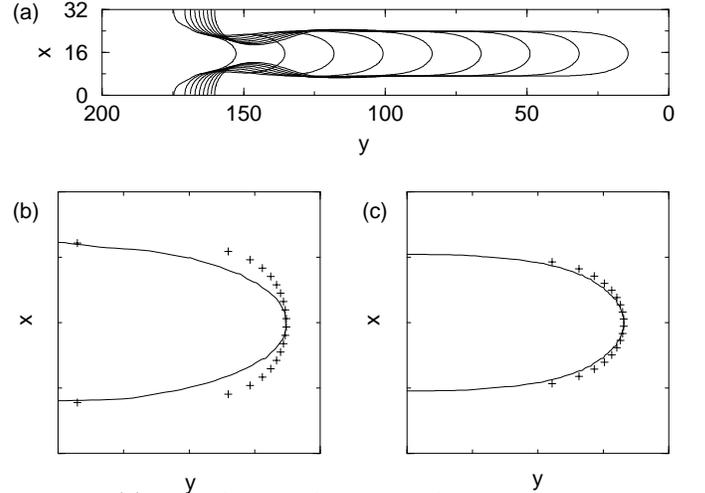}
\caption{
(a) Evolution of a single finger in a channel, plotted at time 
intervals of $750$, for $\alpha=0.035$.
(b) and (c) Numerical results (lines) and Saffman-Taylor solution (symbols)
are presented for two values of $\alpha$ that lead to two different
finger widths (b) $\lambda=0.61$ and (c) $\lambda=0.53$ (c).
}
\label{fig4}
\end{center}
\end{figure}

The width of the steady-state finger is expected to go to
one half of the channel width as the velocity of the finger tip
increases \cite{mclean,pitts}. To better explore this situation,
we have considered a narrow channel of width $L_x=32$,
prepared with an initial condition
that gives a single finger. 
We have analysed the temporal evolution of the finger
for different tip velocities, corresponding to different values
of the parameter $\alpha$. In agreement with known results,
higher velocities led to narrower fingers
An example of the interface evolution is shown in Fig. \ref{fig4}a.
In Figs. \ref{fig4}b and \ref{fig4}c we compare the finger shapes obtained 
numerically for two different values of $\alpha$ with the theoretical shape 
for the Saffman-Taylor finger \cite{st}:
\begin{equation}
y=y_{tip}-\frac{L_x(1-\lambda)}{2 \pi}\ln \left[ \frac{1}{2}
\left (1+\cos\frac{\pi (2 x - L_x)}{\lambda L_x}\right )\right] .
\label{ST}
\end{equation}
being $\lambda$ the ratio of the width of the finger to the width of the
channel.
To determine $\lambda$ from our numerical results, we have evaluated
the average width of the finger throughout the evolution,
in a strip of thickness $e=4$ placed
at distance $40$ from the tip.
For high enough tip velocities, 
our numerical results are in agreement with the Saffman-Taylor solution 
since they correspond to values of $\lambda$ closed to
$1/2$ (Fig. \ref{fig4}c), where surface tension effects are negligible.   
Also, the expected deviation from the Saffman-Taylor solution is
observed for wider fingers in qualitative agreement with
reference \cite{mclean}.
 
\begin{figure}
\begin{center}
\def\epsfsize#1#2{0.50\textwidth}
\leavevmode
\epsffile{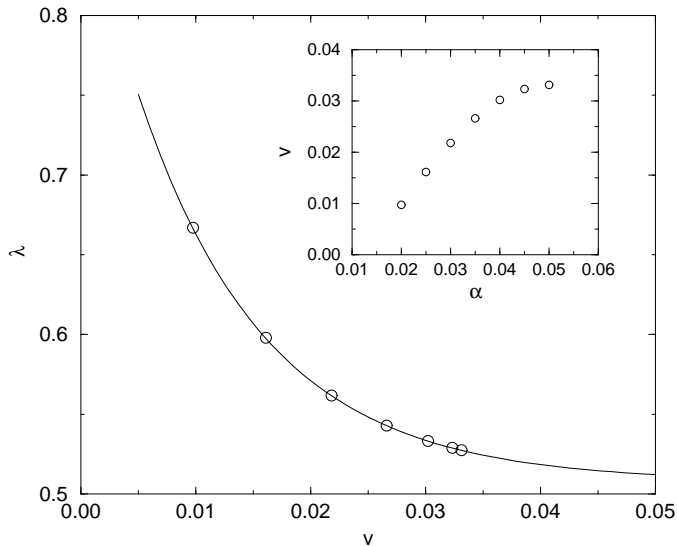}
\caption{
Finger-width $\lambda$ versus 
velocity $v$.
Solid line is a guide to the eye.
Inset shows the dependence of $v$ on the parameter $\alpha$.
}
\label{fig5}
\end{center}
\end{figure}

We have measured the finger 
width and the finger-tip velocity $v$ for different values of the
parameter $\alpha$. The results, shown in 
Fig. \ref{fig5}, are in qualitative agreement with
experimental results of
Pitts \cite{pitts} and Saffman-Taylor \cite{st} 
and with the numerical results of McLean and Saffman 
\cite{mclean}.
We observe that $\lambda$ tends 
to one half the channel width as the velocity increases 
\cite{tabe}.

\section*{Conclusions}

A one-sided phase-field model to describe the dynamic
evolution of unstable interfaces for Hele-Shaw flows in the 
high-viscosity-contrast regime has been proposed. 
The mesoscopic model contains a field equation 
for a conserved order parameter (Model B of Ginzburg-Landau 
phenomenology) and a boundary condition that drives the interface 
out of equilibrium.
An asymptotic expansion to
derive the macroscopic equations has been performed.  
The phase-field model has been numerically integrated and we have 
analysed different stages of the dynamics.
We observe how from a random perturbation to the interface,
fingers develop. Modes grow and compete dynamically and the competition
ends in a single steady-state finger.
The width of this finger goes to one half of the channel width
as the velocity increases. This is in agreement with experiments
and the existent theory.
We have verify that the shape of the finger tip is in good agreement
with the parametric solution of Saffman and Taylor when the
finger width is close to one half of the channel width.
Also for larger width the shape is in qualitative agreement 
with the fingers found by Mc Lean and Saffman.
We believe that our model could be a useful tool to study
situations that cannot be easily tackled with traditional methods,
like integro-differential equations, such as the effect introduced
by quenched disorder.

\section*{Acknowledgements} 
A.H.M. and A.M.L. acknowledge financial support of the Direcci\'on General 
de Ense\~nanza Superior (Spain) under projects BFM2000-0628-C03-01 and 
BFM2000-0624-C03-02, respectively. E.C.P. acknowledges financial support
from PAPIIT through grant IN117802-02.

\end{multicols}

\begin{references}

\bibitem{general}
{\it Branching in Nature}, V. Fleury, J.-F Gouyet and  M. L\'eonetti Eds. 
(Springer-Verlag, Berlin, (2001)); E. Ben-Jacob and H. Levine, 
Adv. Phys. {\bf 49},
395 (2000); J.P. Gollub and J.S. Langer, Rev.Mod. Phys. {\bf 71}, S396 (1999);
{\it Solids far from Equilibrium}, C. Godr\`eche Ed.
(Cambridge University Press,
Cambridge, (1992)); P. Pelc\'e, {\it Dynamics of Curved Fronts} (Academic, New
York, (1988) 
\bibitem{st}
P.G. Saffman and G.I. Taylor, Proc. R. Soc. London A, {\bf 245}, 312 (1958);
D. Bensimon, L. Kadanoff, S. Liang, B.I. Shraiman and C. Tang, Rev. Mod. Phys.,
{\bf 58}, 977 (1986)
\bibitem{mclean}
J.~W. McLean and P.~G. Saffman,
J. Fluid Mech., {\bf 102}, 455, (1981)
\bibitem{tanveer}
M. Siegel and S. Tanveer, Phys. Rev. Lett. {\bf 76}, 419 (1996).
\bibitem{aref}
G. Tryggvason and H. Aref, J. Fluid Mech, {\bf 136}, 1 (1983); 
{\bf 154}, 287 (1985) 
\bibitem{integro}
D. Jasnow and J. Vinals, Phys. Rev. A {\bf 40}, 3864 (1989); {\bf 41}, 6910 (1990);
J. Casademunt, D. Jasnow and A. Hern\'andez-Machado, Int. J. Mod. Phys. B {\bf 6}
1647 (1992); E. Paun\'e, M. Siegel and J. Casademunt, Phys. Rev. E {\bf 66},
046205 (2002)
\bibitem{hou}
T.Y. Hou, J.S. Lowengrub and M.J. Shelley, J. Comp. Phys. {\bf 114}, 312
(1994). 
\bibitem{phase}
J.S. Langer, in {\it Directions in Condensed Matter Physics}, p.165,
Eds. G. Grinstein and G. Mazenko, World Scientific (1986);    
R. Kobayashi, Physica D, {\bf 63}, 410 (1993);
G. McFadden, A. Wheeler, R. Braun, S. Coriell and R. Sekerka,
 Phys. Rev. E, {\bf 48}, 2016 (1993);
K.R. Elder, M. Grant, N. Provatas and J.M. Kosterlitz, Phys. Rev. E {\bf 64}, 
021604 (2001).
\bibitem{folch}
R. Folch, J. Casademunt, A. Hern\'andez-Machado and L. Ram\'{\i}rez-Piscina,
Phys. Rev. E{\bf 60}, 1724 (1999); ibid. {\bf 60}, 1734 (1999) 
\bibitem{gonzalez}
R. Gonzalez-Cinca, R. Folch, R. Benitez, L. Ramirez-Piscina,
J. Casademunt and A. Hernandez-Machado,
cond-mat/0305058, in {\it Advances in Condensed Matter and Statistical
Mechanics}, Ed. E. Korucheva and R. Cuerno, Nova Science Publishers.
\bibitem{machado}
A. Hern\'andez-Machado, J. Soriano, A.M. Lacasta, M.A. Rodr\'{\i}guez, L. Ram\'{\i}rez-Piscina
and J. Ort\'{\i}n. Europhys. Lett. {\bf 55}, 194 (2001).
\bibitem{misbah}
T. Biben and C. Misbah, Phys. Rev. E {\bf 67}, 031908 (2003).
\bibitem{lowengrub}
H.G. Lee, J.S. Lowengrub and J. Goodman, Phys. Fluids {\bf 14}, 492 (2002);
Phys. Fluids {\bf 14}, 514 (2002).
\bibitem{levine}
J. Kockelkoren, H. Levine and W.J. Rappel, cond-mat/0305577.
\bibitem{karma}
A. Karma, Phys. Rev. Lett., {\bf 87}, 115701 (2001)
\bibitem{hohenberg}
P. C. Hohenberg and B. I. Halperin, Rev. Mod. Phys., {\bf 49},
435 (1977)  
\bibitem{almgren}
R. Almgren, W.-S. Dai and V. Hakim, Phys. Rev. Lett. {\bf 71}, 3461 (1993)
\bibitem{guo}
J.L. Mozos and H. Guo, Europhys. Lett., {\bf 32}, 61 (1995)
\bibitem{foot}
An
asymptotic expansion was presented in Ref.\cite{mozos} within the context of
driven diffusive systems. In this case the destabilization of the
interface has a different origin related to the presence of an external
field and a two-sided symmetric model is considered.
\bibitem{mozos}
C. Yeung, J.L. Mozos, A. Hern\'andez-Machado and D. Jasnow, J. Stat. Phys. 
{\bf 70}, 1149 (1993)
\bibitem{pitts}
E. Pitts, J. Fluid Mech., {\bf 97}, 53, (1980)
\bibitem{tabe}
In our model we have not included the film effect observed by 
P. Tabeling, P.G. Zocchi and A. Libchaber, J. Fluid Mech., {\bf 177}, 67, (1987)   



\end{references}
\end{document}